\documentstyle[ltwol]{article}

\input my_psfig.sty

\def\ra{\rightarrow}

\def\be{\begin{equation}}
\def\ee{\end{equation}}
\def\bea{\begin{eqnarray}}
\def\eea{\end{eqnarray}}

\begin{document}


\newcommand{\cm}{\rm \,cm}
\newcommand{\mm}{\rm \,mm}
\newcommand{\m}{\rm \,m}
\newcommand{\nm}{\rm \,nm}
\newcommand{\minu}{\rm \,min}
\newcommand{\Km}{\rm \,Km}
\newcommand{\e}{\rm e}
\newcommand{\s}{\rm \,s}
\newcommand{\ms}{\rm \,ms}
\newcommand{\sr}{\rm \,sr}
\newcommand{\ns}{\rm \,ns}
\newcommand{\mg}{\rm \,mg}
\newcommand{\lit}{\rm \,l}
\newcommand{\g}{\rm \,g}
\newcommand{\MeV}{\rm \,MeV}
\newcommand{\eV}{\rm \,eV}
\newcommand{\KeV}{\rm \,KeV}
\newcommand{\GeV}{\rm \,GeV}
\newcommand{\TeV}{\rm \,TeV}
\newcommand{\erg}{\rm \,erg}
\newcommand{\Kpc}{\rm \,Kpc}
\newcommand{\Mpc}{\rm \,Mpc}
\newcommand{\kHz}{\rm \,kHz}
\newcommand{\mHz}{\rm \,mHz}
\newcommand{\MHz}{\rm \,MHz}
\newcommand{\Hz}{\rm \,Hz}
\newcommand{\p}{\rm p}
\newcommand{\n}{\rm n}
\newcommand{\deu}{\rm d}
\newcommand{\nue}{$\nu_e$}
\newcommand{\anue}{$\bar{\nu}_e$}
\newcommand{\numu}{$\nu_{\mu}$}
\newcommand{\anumu}{$\bar{\nu}_\mu$}
\newcommand{\nutau}{$\nu_{\tau}$}
\newcommand{\anutau}{$\bar{\nu}_\tau$}
\newcommand{\nux}{$\nu_x$}
\newcommand{\anux}{$\bar{\nu}_x$}
\newcommand{\lsim}{\lower .5ex\hbox{$\buildrel < \over {\sim}$}}
\newcommand{\gsim}{\lower .5ex\hbox{$\buildrel > \over {\sim}$}}
\newcommand{\system}[1]{\left\{\matrix{#1}\right.}
\newcommand{\displayfrac}[2]{\frac{\displaystyle
 #1}{\displaystyle #2}} 
\newcommand{\diff}{{\rm\,d}}

\title{Search for Rare Particles with the MACRO Detector}

\author{Fabrizio Cei for the MACRO Collaboration}

\address{University of Michigan, Department of Physics, 2071 Randall 
Laboratory, 500 East University, Ann Arbor, MI 48109-1120, USA\\E-mail: 
cei@pooh.physics.lsa.umich.edu}   


\twocolumn[\maketitle\abstracts{We present the results of the search for rare 
particles (magnetic monopoles, nuclearites, WIMPs and LIPs) with the 
MACRO detector. For magnetic monopoles (the main goal of the experiment) 
our limit is $\sim 0.4$ times the Parker bound for $10^{-4} \le \beta 
\le 10^{-1}$.}]

\section{Introduction}

MACRO~\cite{MACR88,MACR93} \footnote{For the 
MACRO author list see the D. Michael paper in these proceedings}
at the Gran Sasso Laboratory is a large area 
underground detector devoted to the search for rare events in the 
cosmic radiation. It is optimized to search for GUT magnetic monopoles (MM), 
but can also perform many observations relevant to astrophysics, nuclear, 
particle and cosmic ray physics. The main MACRO physics items are the study of 
atmospheric neutrinos and oscillations, the search for MMs, 
the study of the high energy underground muons, the primary cosmic ray 
composition, the measurement of the muon residual energy spectrum, the search 
for low energy stellar collapse neutrinos and the high energy neutrino 
astronomy. 

Here we present the results of the search for rare particles: magnetic 
monopoles and nuclearites, weakly interacting massive particles and lightly 
ionizing particles. All these searches gave (until now) null results, setting 
significant limits on the fluxes of these rare particles. 

\section{The MACRO detector}

The MACRO detector consists of six supermodules (total sizes $77 \times 12 
\times 9~{\rm m}^3$), each one divided in a lower and an upper part. The lower 
part is made by ten horizontal planes of limited streamer tubes, interleaved 
with seven rock absorber layers, and two liquid scintillation counter layers on 
the top and bottom. The lateral walls are closed by four \lq\lq vertical 
detectors\rq\rq, formed by a liquid scintillator layer sandwiched between 
two sets of streamer tubes (three planes each). The upper part is made by two 
\lq\lq vertical detectors\rq\rq~on the East and West faces and by a roof with 
one layer of scintillators sandwiched between two planes 
of tubes; this part is left open on the North and South faces to house the 
electronics. A nuclear track detector~\cite{MACR94b} is located horizontally 
in the middle of the lower part and vertically on the East and North walls. 

The scintillation counters are equipped with specific triggers for rare 
particles, muons and stellar gravitational collapse neutrinos and by 
$200~{\MHz}$ WFDs. The streamer tubes are read by $8$-channel cards which 
discriminate the signals and send the analog information (time development 
and total charge) to an ADC/TDC system; the discriminated signals form two 
different chains of TTL pulses, which are the inputs for the streamer 
tube Fast and Slow Particle Triggers.

\section{Magnetic Monopoles}

Massive ($M_{M} \sim 10^{17}~GeV/c^2$) magnetic monopoles arise spontaneously 
in Grand Unified Theories (GUTs)~\cite{Pres79} of electroweak and strong 
interactions. Magnetic monopoles of such a large mass cannot be produced 
with accelerators and must be searched in the cosmic radiation. The MACRO 
experiment was designed to be sensitive to monopoles at a flux level well below 
the Parker Bound~\cite{Park82} 
$\Phi_M~\lsim~10^{-15}~{\cm}^{-2} {\s}^{-1} {\sr}^{-1}$ in the monopole 
velocity range $4 \times 10^{-5} < \beta < 1$. The use of three subdetectors 
(liquid scintillators, streamer tubes and nuclear track detector) ensure redundancy 
of information, multiple cross-checks and independent signatures for possible 
monopole candidates. The results reported here are obtained using the various 
subdetectors in a stand-alone and in a combined way. All the limits refer to 
monopoles with unit Dirac magnetic charge ($g = 137/2~e$), catalysis cross 
section $\sigma < 10~{\rm mb}$ (we do not consider the monopole induced 
nucleon decay) and isotropic flux (we consider monopoles with enough kinetic 
energy to traverse the Earth); the last condition sets a $\beta$-dependent 
mass threshold of $\sim 10^{17}~{\rm GeV}$ for $\beta \sim 5 \times 10^{-5}$ 
and lower (down to $10^{10}~{\rm GeV}$) for faster monopoles. 

\vspace{-0.1cm}
\subsection{Searches with the scintillator subdetector}

The energy loss, arrival time and velocity of a particle passing in the 
scintillator system are measured by using the total charge and shape 
of the photomultiplier pulses. 


In the low velocity region ($1.8 \times 10^{-4} < \beta < 3 \times 10^{-3}$) 
we studied the photomultiplier waveforms, looking for the wide, flat and small 
amplitude signals (or long trains of single photoelectrons) expected for a 
monopole. No candidates were found in two independent 
analyses~\cite{MACR94a,MACR97a}; the flux upper limits ($90~\%$ C.L.) 
are $5.6$ and $4.1 \times 10^{-15}~{\cm}^{-2} {\s}^{-1} {\sr}^{-1}$ (curves 
\lq\lq A\rq\rq~and \lq\lq B\rq\rq~in fig.~\ref{macrolim}).

In the medium velocity range ($1.2 \times 10^{-3} < \beta < 10^{-1}$) the 
monopoles are searched using the data collected between October $1989$ and March 
$1998$ by the stellar gravitational collapse trigger PHRASE~\cite{MACR92a}. The 
events selected in this $\beta$ range are rejected since their pulse width is 
smaller than the expected counter crossing time or since the light produced is 
lower than that expected for a monopole~\cite{Ahle83}. The flux upper limit at 
$90~\%$ C.L. is $4.3 \times 10^{-16}~{\cm}^{-2}~{\s}^{-1}~{\sr}^{-1}$ (curve 
\lq\lq D\rq\rq~in fig.~\ref{macrolim}). The technique is fully discussed 
in~\cite{MACR97b}. 

Finally, in the high velocity range ($\beta > 0.1$) the data collected by the 
muon trigger ERP~\cite{MACR92a} are used. All events are rejected since 
the measured energy deposit in two counter layers is much lower than 
the energy loss~\cite{Ahle83} expected for a fast monopole. The $90~\%$ C.L. 
flux upper limit is $4.4 \times 10^{-15}~{\cm}^{-2}~{\s}^{-1}~{\sr}^{-1}$ 
(curve \lq\lq C\rq\rq~in fig.~\ref{macrolim})~\cite{MACR97b}.

\vspace{-0.1cm}
\subsection{Search with the streamer tubes subdetector} 

The MACRO streamer tubes~\cite{MACR93,MACR95b} are filled with a mixture of 
$He$ ($73~\%$) and n-pentane ($27~\%$); the Helium was chosen to allow the 
detection of slow ($\beta~\lsim~10^{-3}$) monopoles by the Drell-Penning 
effect~\cite{Drel83}. The hits on the streamer tubes system and the charge 
collected on each tube provide measurements of track, velocity and energy loss 
of an ionizing particle~\cite{MACR95b}. The data were collected from February 
$1992$ to October $1997$. The monopole analysis is based on the search for 
clean single tracks of well reconstructed velocity; it was checked that the 
trigger selection and analysis procedure are velocity independent. No 
candidated survived; the flux upper limit ($90~\%$ C.L.) is $\Phi < 4.5 \times 
10^{-16}~{\cm}^{-2} {\s}^{-1} {\sr}^{-1}$ for $1.1 \times 10^{-4} < \beta < 
5 \times 10^{-3}$ (curve \lq\lq Streamer\rq\rq~in 
fig.~\ref{macrolim})~\cite{MACR97a,MACR98a}. 

\vspace{-0.1cm}
\subsection{Search with the nuclear track subdetector}

The MACRO nuclear track subdetector is made by three layers of LEXAN and 
three layers of CR$39$; its total surface is $1263~{\m}^2$ and its acceptance 
for fast monopoles is $\approx 7100 {\m}^2 {\sr}$. A calibration of the CR$39$ 
with slow and fast ions showed that its response depends on the restricted 
energy loss only~\cite{MACR94b}. The track-etch subdetector is used in a 
stand-alone or in a \lq\lq triggered\rq\rq~mode by the streamer tubes and the 
scintillator systems. A total surface of $181~{\m}^2$ was etched, with an 
average exposure time of $7.23~{\rm years}$; the flux upper limits ($90~\%$ 
C.L.) are $\sim 0.88$ and $\sim 1.3 \times 
10^{-15}~{\cm}^{-2}~{\s}^{-1}~{\sr}^{-1}$ at $\beta \sim 1$ and $\beta 
\sim 10^{-4}$ respectively (curves \lq\lq CR39\rq\rq~in 
fig.~\ref{macrolim})~\cite{MACR97a,MACR98a}. 

\vspace{-0.1cm}
\subsection{Combined search}

The fast monopoles are expected to release a huge amount of energy by 
ionization/excitation without producing showers for $\beta < 0.99$. A monopole 
search based on the energy deposition only can be seriously affected by a large 
background due to showering cosmic rays, but such a background is efficiently 
rejected by a combined analysis. This analysis uses at the same time the data 
collected by the scintillators and by the streamer tubes, requiring large 
photomultiplier pulses, an isolated single track and a high streamer charge 
per unit path length. Only few ($\sim 5/{\rm year}$) events survive and are 
looked for in the appropriate track-etch sheets. In $667~{\rm days}$ of live 
time 
no candidates were found; the $90~\%$ C.L. flux upper 
limit is $1.5 \times 10^{-15}~{\cm}^{-2}~{\s}^{-1}~{\sr}^{-1}$ for $5 \times 
10^{-3} < \beta < 0.99$ (curve \lq\lq E\rq\rq~in 
fig.~\ref{macrolim})~\cite{MACR97c}. 

\vspace{-0.1cm}
\subsection{Conclusions about magnetic monopoles}

\begin{figure}[htb]
\begin{center}
\mbox{
	\hspace{-1.6cm}\psfig{file=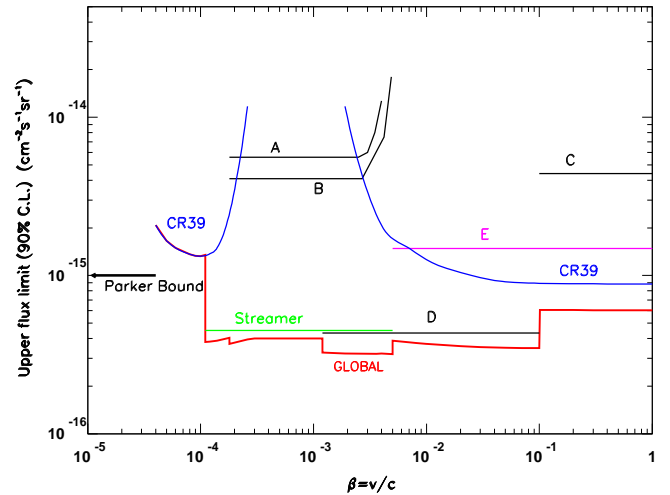,height=3.2in}
     }
\end{center}
\vspace{-1.5cm}
\caption{Magnetic monopole flux upper limits obtained using the various MACRO 
subdetectors.}
\label{macrolim}
\end{figure}

\begin{figure}[htb]
\begin{center}
\vspace{1.5cm}
\mbox{
	\hspace{-1.2cm}\psfig{file=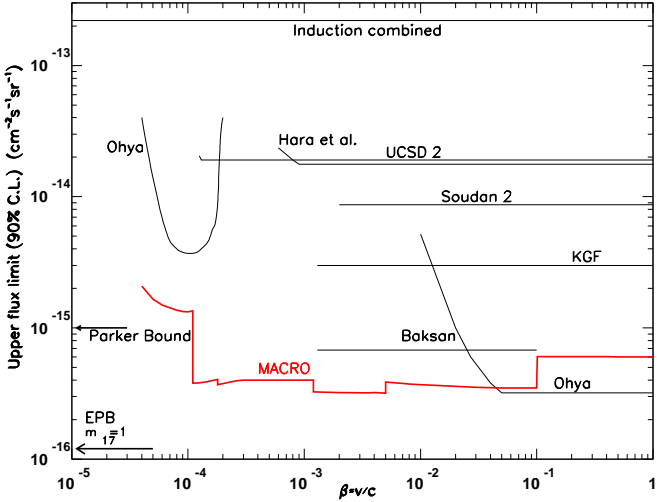,height=3.88in}
    }
\end{center}
\vspace{-3.5cm}
\caption{Magnetic monopole flux upper limits obtained by MACRO and by other 
experiments.} 
\label{worldlim}
\end{figure}

We show in fig.~\ref{macrolim} the flux upper limits obtained using the 
various MACRO subdetectors. Since each subdetector can rule out, within its 
acceptance and sensitivity, a potential candidate from the others, we obtain a 
global MACRO limit (curve \lq\lq GLOBAL\rq\rq~in fig.~\ref{macrolim}), as an 
\lq\lq OR\rq\rq~combination of the separate results. The prescriptions used 
for this combination are described in~\cite{MACR97a}. 
%
In fig.~\ref{worldlim} we compare the MACRO combined result with that 
obtained by other 
experiments~\cite{Berm90,Buck90,Thro92,Baks90,Orit91,Kola90,Hara90} and with 
the Parker Bound. 
Our limit is at the level of 0.3 times the Parker Bound for $\beta > 
10^{-4}$ and is the best existing for $10^{-4} < \beta < 5 \times 10^{-2}$. 

\section{Nuclearites}

\begin{figure}[htb]
\begin{center}
\mbox{
	\hspace{-0.2cm}\psfig{file=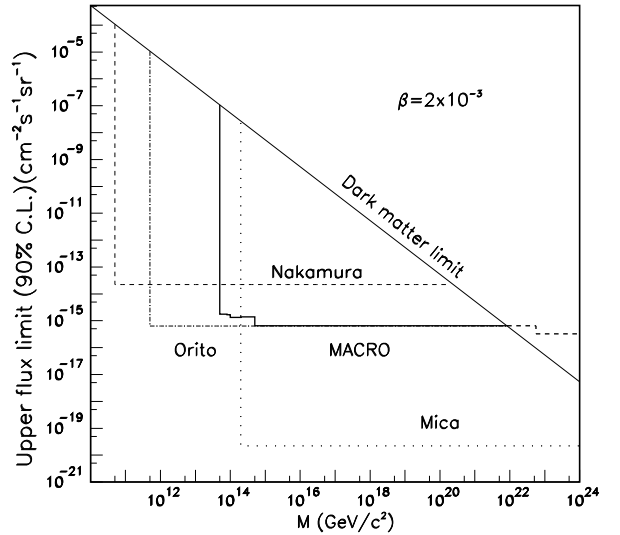,height=3.2in}
    }
\end{center}
\vspace{-0.3cm}
\caption{Nuclearite flux upper limits obtained by MACRO and by other 
experiments; the dark matter bound is also shown.} 
\label{nuclim}
\end{figure}

The results obtained using the liquid scintillator and the nuclear track 
subdetectors can be, at least in part, extrapolated to the search for 
nuclearites~\cite{Witt84}, hypothesized nuggets of strange quark matter 
(the streamer tubes are not sensitive to nuclearites because of the low 
density of the filling gas~\cite{MACR97d}). By studying the mechanism of 
nuclearite energy loss~\cite{Deru84} it was shown that the scintillators are 
sensitive to nuclearites~\cite{MACR92b} down to $\beta~\lsim~10^{-4}$ and the 
CR$39$~\cite{MACR97d} down to $\beta \approx 10^{-5}$. The MACRO flux 
upper limits are $\sim 4 \times 10^{-16}~{\cm}^{-2}~{\s}^{-1}~{\sr}^{-1}$ 
for $M_{N} > 0.1~{\rm g}$ and about $2$ times higher for $M_{N} < 0.1~{\rm 
g}$ (the lighter nuclearites cannot traverse the Earth and then only the 
down-going flux must be considered). Assuming a nuclearite velocity 
at the ground level $\beta = 2 \times 10^{-3}$ we compared our limit with the 
results of other experiments~\cite{Orit91,Naka91,Pri88} and with the dark 
matter bound; our results look competitive in a large range of nuclearite 
masses ($10^{14}~{\GeV}/c^2~\lsim~M_{N}~\lsim~10^{21}~{\GeV}/c^2$) 
(see fig.~\ref{nuclim} \footnote{The limit is prolonged above the Dark 
matter bound to show the transition to an isotropic flux for $M > 0.1~{\rm 
g}~\left( \approx 4.5 \times 10^{22}~{\rm GeV/c^2} \right)$}.)
%
For further details on this search see~\cite{MACR97d}. 

\section{WIMPs}
The {\it Weakly Interacting Massive Particles} (WIMPs) are important 
candidates for the Cold non-baryonic part of the Dark Matter in the 
Universe~\cite{Prima96}. Between the various Cold Dark Matter candidates 
subject to weak interactions one of the most promising is the 
supersymmetric neutralino $\tilde{\chi}$~\cite{Jung96}. 

In supersymmetric theories where the $R$ parity is conserved there exists a 
lightest stable supersymmetric particle (LSP), which is the natural candidate 
for the Dark Matter since its expected density is close to the critical 
one: $\Omega_{LSP} \sim 1$. In many theories the LSP is the neutralino 
$\tilde{\chi}$, the simplest linear combination of the gaugino and higgsino 
eigenstates. The $\tilde{\chi}$ mass depends on the supersymmetric parameters, 
as, for instance, the gaugino and higgsino mass parameters $M_1$, $M_2$ 
and $\mu$ and the ratio of the Higgs doublet vacuum expectation values 
$\tan \beta$. These parameters are constrained by accelerator searches and a 
lower limit on $m_{\tilde{\chi}}$ was set by LEP $2$ data: $m_{\tilde{\chi}} 
> 20 \div 30~{\rm GeV}$~\cite{LEP96}. The search for WIMPs in underground 
detectors can probe complementary regions of the parameter space. 

\vspace{-0.1cm}
\subsection{WIMP searches in MACRO}

The WIMPs are indirectly searched in MACRO by using upward going muons. 
A WIMP intercepting a celestial body can lose its energy and be trapped in 
the core of this body and annihilate with an other WIMP; the decay of the 
annihilation products produces high-energy $\nu$'s which can be detected in 
an underground detector as upward going muons. Then, a statistically 
significant excess of upward going muons from the direction of a celestial body 
can be a hint for a WIMP-WIMP annihilation in the core of 
that body. Some possible traps were proposed and some calculations of the WIMP 
annihilation rate in the Earth and Sun and of the corresponding upward going 
muon fluxes were performed~\cite{Jung96,Bott95,Berg97}. This indirect search 
achieves a better signal to noise ratio for high WIMP masses, since the higher 
the WIMP mass, the more the upgoing muon follows the parent neutrino direction. 
The technique was already used by other experiments~\cite{Baks96,Kam93,IMB87}.  

The upward going muons in MACRO are selected by the time-of-flight 
technique, requiring a track in the streamer tubes and a time of flight 
between the scintillation counter layers consistent with a $\beta \sim 1$
particle from below. Details on the selection criteria can be found 
in~\cite{Mich98}. 

\vspace{-0.1cm}
\subsection{Search for WIMP annihilation in the Earth}

Upward going muons from WIMP annihilation in the Earth core are searched in 
angular cones ($3 \div 30^{\circ}$ wide) around the vertical. We used 
$517$ events, collected in $3.1$ years of live time, requiring a minimum 
crossing of $200~{\rm g}~{\rm cm}^{-2}$ of rock absorber. After background 
subtraction the number of selected events is $487~\pm~22_{stat}$, to be 
compared with $653~\pm~111_{theor}$ expected from a Monte Carlo calculation. 
Most of the deficit lies around the vertical, where the WIMP 
annihilation signal is expected, but also where the efficiency and acceptance 
of the apparatus are best known. To set a conservative limit on the WIMP flux 
we assumed that the number of measured and expected events are equal and 
normalized the expected distribution to the factor $0.85$, corresponding to the 
ratio between measured and expected events for $\theta > 30^{\circ}$. With this 
prescription our limits on the WIMP flux from the Earth range from $0.4$ to 
$2.3 \times 10^{-14}~{\rm cm}^{-2}~{\rm s}^{-1}$ for angular windows from $3$ 
to $30^{\circ}$. 

\vspace{-0.1cm}
\subsection{Search for WIMP annihilation in the Sun}

In the search for WIMPs from the Sun, since the background for moving sources  
is lower than for steady sources, we used an enlarged sample ($762$ events)
which includes also muons partially contained in the apparatus (produced by 
neutrino interactions in the absorber in the MACRO lower part) and muons 
crossing $< 200~{\rm g}~{\rm cm}^{-2}$ of rock absorber. The simulation was 
obtained by the data themselves to include properly the effects of the 
partially contained events and the arrival times were extracted randomly during 
the whole measurement time to take into account possible drifts of detection 
efficiency. The angular distribution of the upward going muon events do not 
show any significant excess around the Sun direction. Then, we set upper limits 
on the WIMP flux from the Sun ranging from $1.7$ to $6.0 \times 10^{-14}~{\rm 
cm}^{-2}~{\rm s}^{-1}$ in the angular window $3 \div 30^{\circ}$. 

\vspace{-0.1cm}
\subsection{Limits on the WIMP fluxes and $\tilde{\chi}$ mass from the 
angular distributions}

\begin{figure}[htb]
\begin{center}
\mbox{
	\hspace{-1.2cm}\psfig{file=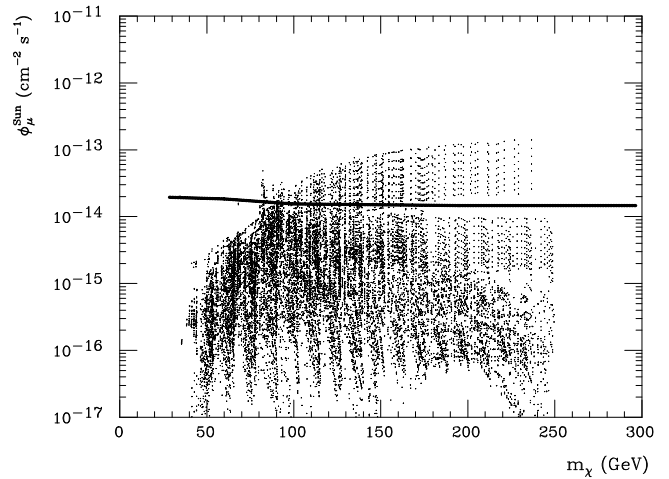,height=4.1in}
     }
\end{center}
\vspace{-2.1cm}
\caption{Upward going muon flux from the Sun vs $\m_{\tilde{\chi}}$ computed 
in$^{31}$. The solid line is the $90\,\%$ C.L. MACRO flux limit.}
\label{WIMPSun}
\end{figure}

As already stated, the angle between the upward going muon and the neutrino 
from the WIMP annihilation depends mainly on the WIMP mass. We performed a full 
Monte Carlo calculation of the expected angle between the upward going muon and 
the Earth or Sun directions for neutralino masses from $60~{\rm GeV}$ to 
$1000~{\rm GeV}$. This calculation uses the shape of the upward going muon 
signals from $\tilde{\chi}-\tilde{\chi}$ annihilation in the Earth and the Sun 
computed in~\cite{Bott95}, propagates the muons through the rock to the 
apparatus (with the cross sections given in~\cite{Lipa95}) and includes the 
angular smearing produced by the experimental resolution. As expected, the 
angle becomes narrower when $m_{\tilde{\chi}}$ increases, but also in the less 
favourable case more than $90\,\%$ of the signal is contained in an angular 
cone $\theta < 15^{\circ}$ around the selected direction. For each 
$\m_{\tilde{\chi}}$ we determine the cone which collects the $90\,\%$ of the 
expected signal and define a corresponding $90\,\%$ C.L. limit on the upward 
going muon flux from $\tilde{\chi}-\tilde{\chi}$ annihilation. These limits 
for the Sun are shown in fig.~\ref{WIMPSun} and compared with the fluxes 
computed in~\cite{Bott95}, varying some supersymmetric parameters 
($M_{1}, \mu, \tan \beta$ etc.). 
\vspace{-1.2cm}
Fig.~\ref{WIMPSun} shows that, thanks to the improved statistics and 
exposure, our data start to constrain the supersymmetric theoretical 
models~\cite{MACR98b}. 

\section{LIPs}
Fractionally charged particles have been actively searched for since many 
years, but without success; the detection of such particles would be a 
proof of their existence and/or of a lack of the confinement hypothesis 
under some circumstances. The family of the possible fractionally charged 
particles includes the quarks~\cite{GeMann64} (with charge $\left| e 
\right|/3$ and $2/3~\left| e \right|$) and a variety of other particles 
predicted by Grand Unified Theories, with charges ranging from $1/5~\left| e 
\right|$ up to $2/3~\left| e \right|$~\cite{Framp82,Bar83,Dong83}. Particles 
with fractional charge deposit less energy than particle with unit charge 
since the energy loss is proportional to the square of the charge; such 
particles are called {\it Lightly Ionizing Particles} (LIPs). 

\begin{figure}[htb]
\begin{center}
\mbox{
	\hspace{-0.25cm}\psfig{file=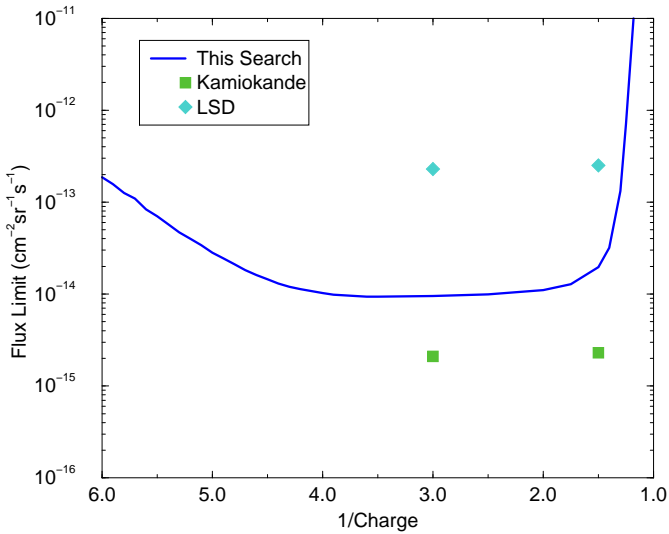,height=2.8in}
    }
\end{center}
\caption{$90\,\%$~C.L. flux upper limits for LIP set by MACRO (solid line) and 
by other experiments as a function of the LIP charge.} 
\label{LIPlim}
\end{figure}

A specialized trigger was developed in MACRO for the LIP search; this trigger 
uses the low energy events collected by the PHRASE system and performs a 
four-fold coincidence between the signals coming from three counters (each one 
in a different scintillator layer) and from the streamer tubes. The LIP 
trigger is sensitive down to $\left| Q/e \right| = 1/5$; the corresponding 
energy loss in the MACRO counters is $\Delta E \approx 1.6~{\MeV}$. The LIP 
trigger gives the stop to the WFD system; the WFD data provide high quality 
measurements of the energy and timing of the events. For each LIP event, the 
maximum energy loss in the three scintillator layers is computed and compared 
with that expected for a fractionally charged particle. The efficiency as a 
function of the charge is obtained by a Monte Carlo simulation which takes into 
account the trigger behaviour close to the threshold and the cosmic muon 
background. No candidate survived out of $1.2$ 
million triggers; the $90 \, \%$~C. L. upper limit for an isotropic flux is 
$\Phi~\le 9.2 \times 10^{-15}~{\cm}^{-2} {\s}^{-1} {\sr}^{-1}$. This limit is 
shown in fig.~\ref{LIPlim} (solid line) and compared with that set by other 
collaborations~\cite{LSD94,Kam91}. 

We stress the fact that our experiment is the first one 
to be sensitive down to $\left| Q/e \right| = 1/5$ and that we can perform a 
high quality search thanks to the combined signature provided by the streamer 
tubes and the scintillators: only few events over $1.2$ millions needed a hand 
scanning (instead of one event over few thousands in Kamiokande). This search 
is discussed in detail in~\cite{Walt97}.




\section*{References}
\begin{thebibliography}{99}

\bibitem{MACR88} MACRO Coll. NIM {\bf A264} (1988) 18
\bibitem{MACR93} MACRO Coll. NIM {\bf A324} (1993) 337
\bibitem{MACR94b} MACRO Coll. LNGS {\bf 94/115} (1994)
\bibitem{Pres79} J. Preskill Phys. Rev. Lett. {\bf 43} (1979) 1365
\bibitem{Park82} E. M. Parker et al. Phys. Rev. {\bf D26} (1982) 1926
\bibitem{MACR94a} MACRO Coll. Phys. Rev. Lett. {\bf 72} (1994) 608
\bibitem{MACR97a} MACRO Coll. Phys. Lett. {\bf B406} (1997) 249
\bibitem{MACR92a} MACRO Coll. Astropart. Phys. {\bf 1} (1992) 11
\bibitem{Ahle83} S. Ahlen \& G. Tarl\'{e} Phys. Rev. {\bf D27} (1983) 688 
\bibitem{MACR97b} MACRO Coll. Astropart. Phys. {\bf 6} (1997) 113 
\bibitem{MACR95b} MACRO Coll. Astropart. Phys. {\bf 4} (1995) 33
\bibitem{Drel83} S. Drell et al. Phys. Rev. Lett. {\bf 50} (1983) 644
\bibitem{MACR98a} MACRO Coll. MACRO/PUB {\bf 98/3} (1998)
\bibitem{MACR97c} MACRO Coll. INFN/AE-{\bf 97/19} (1997)
\bibitem{Berm90} S. Bermon et al. Phys. Rev. Lett. {\bf 64} (1990) 839
\bibitem{Buck90} K. N. Buckland et al. Phys. Rev. {\bf D41} (1990) 2726
\bibitem{Thro92} J. L. Thron et al. Phys. Rev. {\bf D46} (1992) 4846
\bibitem{Baks90} Baksan Coll. $21^{st}$ ICRC, Adelaide (1990) 83
\bibitem{Orit91} S. Orito et al. Phys. Rev. Lett. {\bf 66} (1991) 1951
\bibitem{Kola90} Kolar Coll. $21^{st}$ ICRC, Adelaide (1990) 95
\bibitem{Hara90} T. Hara et al. $21^{st}$ ICRC, Adelaide (1990) 79
\bibitem{Witt84} E. Witten Phys. Rev. {\bf D30} (1984) 272
\bibitem{Deru84} A. De R\'{u}jula \& S. Glashow Nature {\bf 312} (1984) 734
\bibitem{Naka91} S. Nakamura et al. Phys. Lett. {\bf B263} (1991) 529
\bibitem{Pri88} P. B. Price Phys. Rev. {\bf D38} (1988) 3813
\bibitem{MACR92b} MACRO Coll. Phys. Rev. Lett. {\bf 69} (1992) 1860
\bibitem{MACR97d} MACRO Coll. INFN/AE-{\bf 97/20} (1997)
\bibitem{Prima96} J. R. Primack DARK'96, Heidelberg (1996)
\bibitem{Jung96} G. Jungman et al. Phys. Rep. {\bf 267} (1996) 1955
\bibitem{LEP96} OPAL Coll. accepted by Z. Phys. {\bf C}; 
ALEPH Coll. Z. Phys. {\bf C72} (1996) 549
\bibitem{Bott95} A. Bottino et al. Astropart. Phys. {\bf 3} (1995) 65
\bibitem{Berg97} L. Bergstr\"{o}m et al. Phys. Rev. {\bf D55} (1997) 1765
\bibitem{Baks96} Baksan Coll. Nucl. Phys. {\bf B48} (Proc. Suppl.) (1996) 83
\bibitem{Kam93} Kamiokande Coll. Phys. Rev. {\bf D48} (1993) 5505 
\bibitem{IMB87} IMB Coll. Phys. Lett. {\bf B188} (1987) 388
\bibitem{Mich98} D. Michael this proceedings; MACRO Coll. submitted to 
Phys. Lett.
\bibitem{Lipa95} P. Lipari et al. Phys. Lett. {\bf 74} (1995) 4384
\bibitem{MACR98b} MACRO Coll. in preparation
\bibitem{GeMann64} M. Gell-Mann Phys. Lett. {\bf 8} (1964) 214
\bibitem{Framp82} P. H. Frampton \& T. Kephart
Phys. Rev. Lett. {\bf 49} (1982) 1310
\bibitem{Bar83} S. M. Bar et al. Phys. Rev. Lett. {\bf 50} (1983) 317
\bibitem{Dong83} F. xiao Dong et al. Phys. Lett. {\bf 129B} (1983) 405
\bibitem{LSD94} LSD Coll. Astropart. Phys. {\bf 2} (1994) 29
\bibitem{Kam91} Kamiokande Coll. Phys. Rev. {\bf D43} (1991) 2843
\bibitem{Walt97} C. W. Walter Ph. D. thesis, California Institute of 
Technology (1997); MACRO Coll. in preparation

\end{thebibliography}
\end{document}